\documentclass[a4paper,
              ]{jacow}

\makeatletter%                           % test for XeTeX where the sequence is by default eps-> pdf, jpg, png, pdf, ...
\ifboolexpr{bool{xetex}}                 % and the JACoW template provides JACpic2v3.eps and JACpic2v3.jpg which might generates errors
 {\renewcommand{\Gin@extensions}{.pdf,%
                    .png,.jpg,.bmp,.pict,.tif,.psd,.mac,.sga,.tga,.gif,%
                    .eps,.ps,%
                    }}{}
\makeatother

\ifboolexpr{bool{xetex} or bool{luatex}} % test for XeTeX/LuaTeX
 {}                                      % input encoding is utf8 by default
 {\usepackage[utf8]{inputenc}}           % switch to utf8

\usepackage[USenglish]{babel}

\ifboolexpr{bool{jacowbiblatex}}%        % if BibLaTeX is used
 {%
  \addbibresource{jacow-test.bib}
  \addbibresource{biblatex-examples.bib}
 }{}

\begin{document}

\title{Fibre Monitoring System for the Beam Permit Loops at the 
\NoCaseChange{LHC} and Future Evolution of the Beam Interlock System}

\author{C. García-Argos, R. Denz, S. Gabourin, C. Martin, \\
B. Puccio, A. Siemko, CERN, Geneva, Switzerland}

\maketitle

\begin{abstract}
The optical fibres that transmit the beam permit loop signals at the CERN 
accelerator complex are deployed along radiation areas. This may result in 
increased attenuation of the fibres, which reduces the power margin of the 
links. In addition, other events may cause the links not functioning properly 
and result in false dumps, reducing the availability of the accelerator chain 
and affecting physics data taking. In order to evaluate the state of the 
fibres, an out-of-band fibre monitoring system is proposed, working in 
parallel to the actual beam permit loops. The future beam interlock system to 
be deployed during LHC long shutdown 2 will implement online, real-time 
monitoring of the fibres, a feature the current system lacks. Commercial 
off-the-shelf components to implement the optical transceivers are proposed
whenever possible instead of ad-hoc designs.
\end{abstract}

\section{Introduction}

The Beam Interlock System (BIS)~\cite{BIS} of the CERN accelerator chain is 
responsible for transmitting the beam permit along the Large Hadron 
Collider~(LHC), the Super Proton Synchrotron~(SPS), the transfer lines and the
PS Booster.

The beam permit loop signals are two different square signals with frequencies
$9.375$ (loop A) and $8.375~{\rm MHz}$ (loop B), sent in opposite 
directions. 
These beam permit loop signals are transmitted over single-mode
optic fibre.

\begin{figure}[!htb]
   \centering
   \includegraphics*[width=80mm]{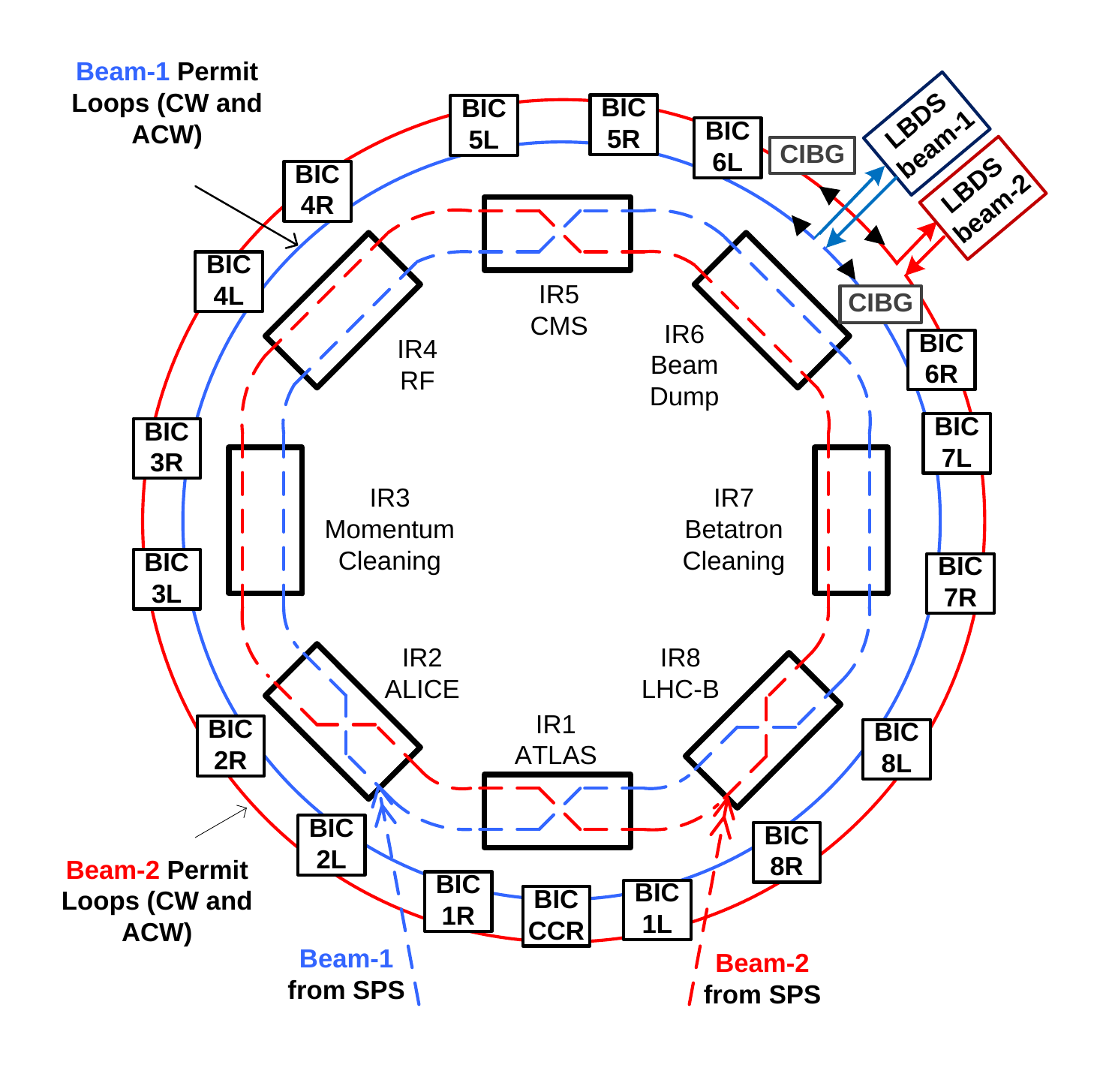}
   \caption{LHC beam permit loops.}
   \label{fig:bpl}
\end{figure}

Fig.~\ref{fig:bpl} shows the topology of the beam permit loops at the LHC. 
There are two signals for each of the two beams, one transmitted clockwise 
and the other anti-clockwise. 

There are seventeen Beam Interlock Controllers (BIC) in the LHC, two at each 
LHC point, named with the point number and side (left or right), and one at the 
CERN Control Centre~(CCC), named CCR. The two generators of the beam
permit signal,  named CIBG, are installed in point 6, where the dump system is 
located.

In the SPS, there is a similar architecture, with one BIC per point, and two
loops for one beam. Injection and extraction lines also have their own beam
permit loops.

The controllers receive the users inputs, coming from the user systems. These 
inputs are connected with a logical AND inside the controller, resulting in
the local permit at the BIC. If the local permit is true, then the BIC 
re-transmits the beam permit signal to the next BIC.

A total of 12 fibres are deployed at each controller: one for each incoming
signal and one for each outgoing signal, for a total of eight active fibres.
There are wo spare fibres to each of the neighbour controllers. The distance 
between controllers is varied, as short as a metre and as long as 6 kilometres.

\section{Out-of-band fibre monitoring}

\subsection{Motivation}

The beam permit loops and the implementation of the optical links are described
in~\cite{BeamPermitOptics}. The Controls Interlocks Beam Optical (CIBO) board
designed at CERN uses an ELED single-mode transmitter and a PIN diode receiver
to implement the optical transceiver for the beam permit signals. 

The working wavelength of the loops is $1310~{\rm nm}$, and the receiver has
a sensitive response in the range $(900, 1700)~{\rm nm}$. The output power of
the transmitter is typically between $-25$ and $-15~{\rm dBm}$ and the CIBO 
board is designed to deliver around $-19~{\rm dBm}$.

The G.652~\cite{G.652} type optical fibres have a maximum length of $6~{\rm km}$, 
with an initial worst case attenuation of $0.5~{\rm dB/km}$ at $1310~{\rm nm}$.
A number of false dumps have occurred that may have been caused by increased
attenuation in the fibres, what drives the need for a monitoring system to 
evaluate their performance during operation.

Radiation can both create point defects in 
the silica and activate already existing defects, causing radiation induced 
attenuation~(RIA)~\cite{RadDamage}. Therefore, a system to monitor the fibres
attenuation is advisable.

Such a monitoring system for the LHC and SPS fibres must not interfere with the
Beam Permit Loop signals, as they are critical to the accelerator operation.
It would be convenient to have measurements of the fibres attenuation over time,
and the capability to monitor both spare and active fibres.

The existing beam permit loop signals cannot be measured directly due to the
tight power margins involved. The use of a tap with sufficient coupling ratio 
that allows for the measurement of the optical power is discouraged because
of the high extra losses on the links.

As an alternative, the transmission of a separate optical signal over the 
existing fibres
is proposed. The use of Wavelength Division Multiplexing enables the 
transmission of multiple signals over the same fibre.

A Wavelength Division Multiplexer~(WDM) is a passive and bi-directional device, which
has one common port, in which multiple wavelengths can be transmitted at the 
same time, and two (or more) ports which only allow one wavelength. 

\subsection{Monitoring System}

A large wavelength separation of the monitoring signal with respect to the beam
permit loop signals was evaluated. The chosen wavelength is $1550~{\rm nm}$, 
which is in a separate window from the beam permit loop transmitter. 
Nevertheless, the receiver is sensitive to this window, an effect that has to
be evaluated in order to verify that the beam permit loops are not disturbed
by the monitoring signal.

Standard commercial components for the system were chosen in order to ease
development and deployment. The transceivers are Standard Small Form-factor 
Pluggable~(SFP)~\cite{SFPMSA}. These typically implement a diagnostics information 
interface~\cite{SFF8472} which allows for the measurement of transmitted and
received power. 

Commercial network switches are used to house, power and access the 
monitoring information on the SFPs. A computer is used to connect to the 
switches, retrieve the information and process it. 

In order to be able to separate the optical signals, commercial WDMs have 
been evaluated and a system has been proposed and 
tested. The isolation
provided by these devices is tight for $1550~{\rm nm}$ systems, as these are
high power devices, in the order of $0~{\rm dBm}$, two orders of magnitude 
higher than the beam permit loop signals. Nevertheless, disabling the ports
on the switch eliminates the interference from the SFP transceiver to the
beam permit loops receivers, while still permitting the power measurements.

The transmission of the beam permit loop signals through WDMs does not cause
any significant signal degradation.

\begin{figure}[!htb]
   \centering
   \includegraphics*[width=80mm]{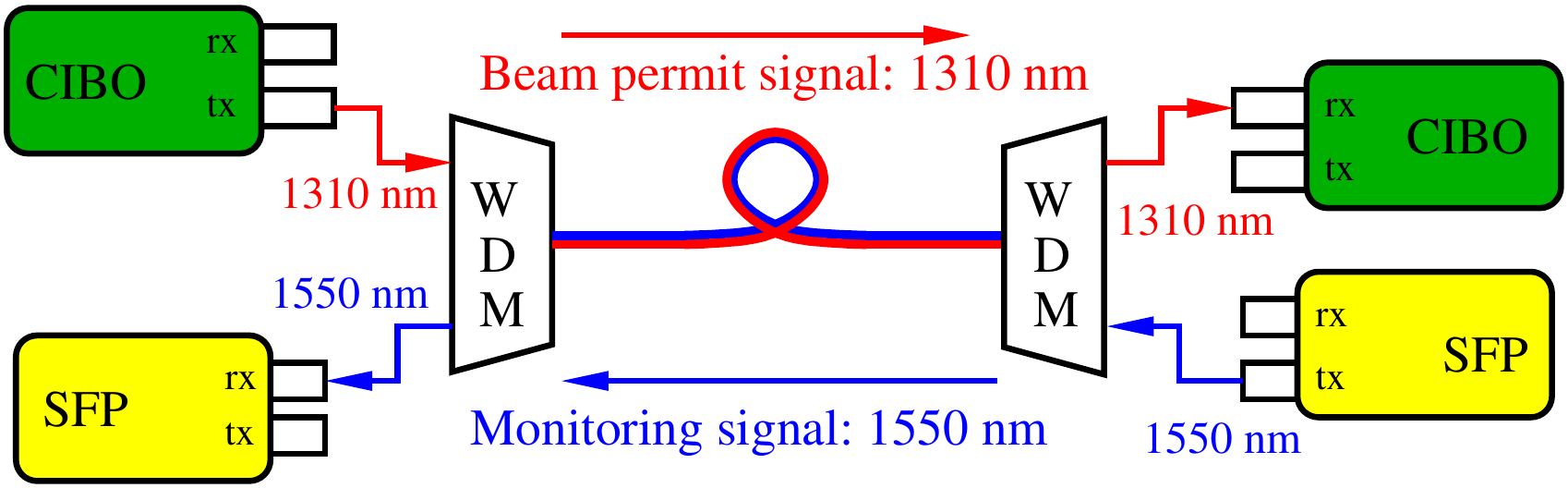}
   \caption{Monitoring system using WDMs and SFPs to transmit a monitoring
   signal in parallel to the beam permit signal.}
   \label{fig:monitorsystem}
\end{figure}

A proposed topology to monitor a link is shown in Fig.~\ref{fig:monitorsystem},
where the two points of one beam permit loop link are shown. The monitoring
signal is sent in the opposite direction from the beam permit, to take 
advantage of a higher isolation between the two input ports of the WDM.

Due to the large power margin of the SFP transceivers, attenuators are also
used in this topology to reduce the power in addition to the WDM isolation. 

A first test system was installed in the SPS spare fibres, between points BA4
and BA6. A BIC was installed and configured in BA5, together with two switches
that held the SFP transceivers to monitor the attenuation. The BIC was set to
latch mode and the whole system ran for 30 days, without observing any losses
of the beam permit. Attenuators were connected at the transmitter outputs of
the SFP transceivers, with an attenuation of $10~{\rm dB}$.

\begin{figure}[!htb]
   \centering
   \includegraphics*[width=80mm]{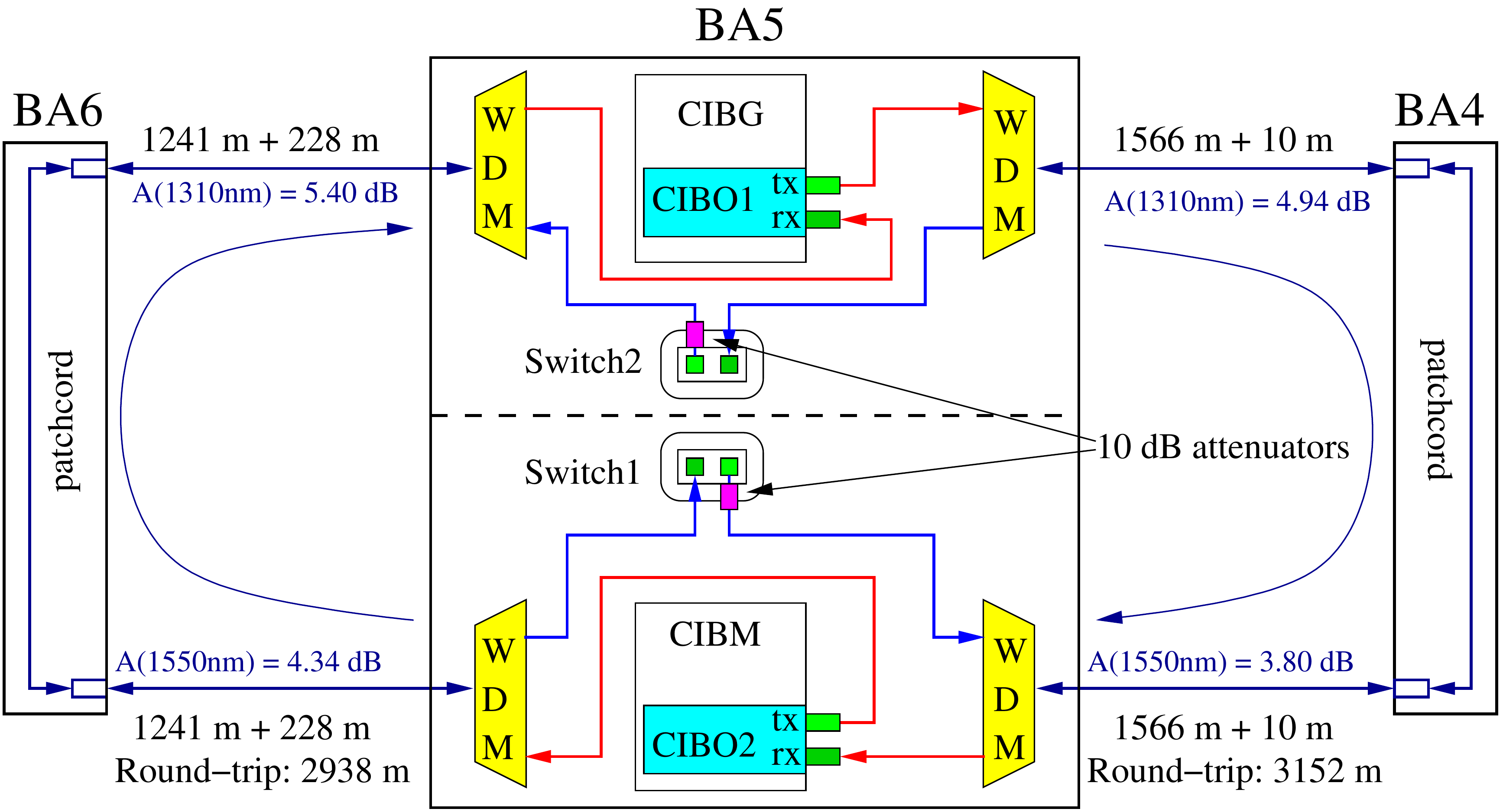}
   \caption{Test setup in the SPS with two switches, WDMs and a Beam Permit Loop
   emulating two locations using three SPS sites.}
   \label{fig:spstest}
\end{figure}

A drawing of the test setup is shown in Fig.~\ref{fig:spstest}. The topology
effectively emulates two separate locations, with one CIBO transceiver at each
one, and the monitoring system connected to two fibres.

\subsection{LHC Fibre Monitoring}

The system is deployed at the LHC, measuring a selection of spare fibres around
the ring. In order to avoid any interference with the system due to the 
connection of the WDMs, no live fibres are monitored. Since the measured fibres 
are spares and there is no beam permit signal on those, no WDMs are required.
The information from this setup is nonetheless useful to track whether the 
fibres are being affected by radiation while the LHC is in operation.

In order to limit the number of switches, three locations were 
selected to install the switches: points 1, 3 and 7. Loopback connections are
in place so the fibres are measured from each switch alone. Bypass connections 
are used at the shortest paths at the same LHC point to go to the next LHC 
point.

The monitored fibres are (all returning to the first point):

\begin{itemize}
 \item US15 to UA23 (point 1-point 2).
 \item US15 to CCR (point 1-CCR).
 \item US15 to UA87 (point 1-point 8).
 \item SR3 to UA43 (point 3-point 4).
 \item SR3 to UA27, with a bypass at UJ33 (point 3-point 2).
 \item SR7 to UA67 (point 7 to point 6).
 \item SR7 to UA83, with a bypass at TZ76 (point 7-point 8).
\end{itemize}

\begin{figure}[!htb]
   \centering
   \includegraphics*[width=82mm]{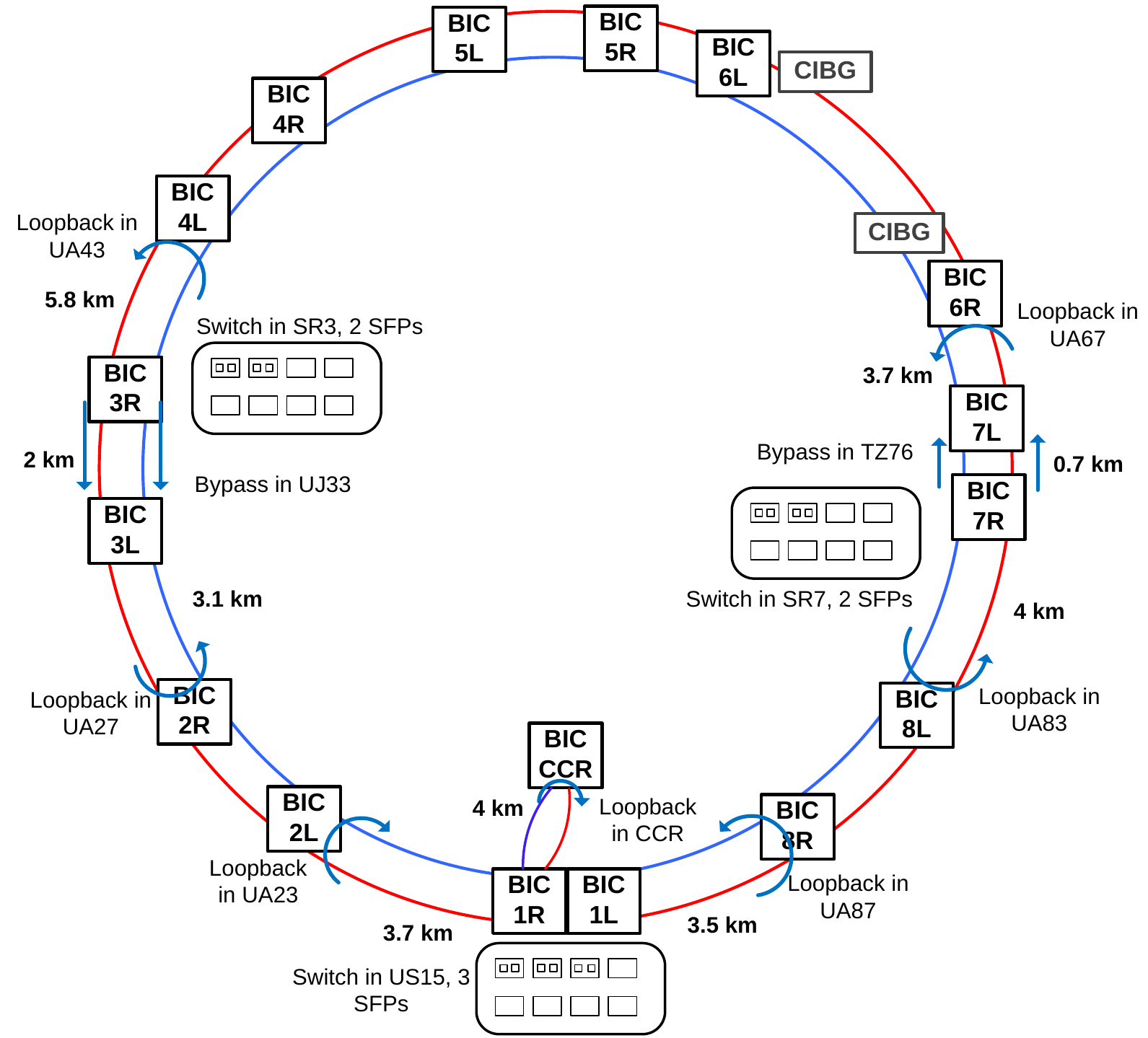}
   \caption{Fibre monitoring system on LHC spare fibres. Bypasses and loopback
   connections are shown, as well as the location of the switches holding the 
   SFPs.}
   \label{fig:sparemonitor}
\end{figure}

The topology of the deployed monitoring system is shown in 
Fig.~\ref{fig:sparemonitor}. The choice of the fibres to be 
monitored was made so they are close to high radiation areas, such as the 
collimators in points 3 and 7.

\section{Upgrade of the Beam Interlock System}

An upgrade of the beam interlock system should be ready, at the earliest,
for the long shutdown 2, starting in 2018. 

Some of the components used in the design of the current BIS
are close to becoming deprecated by the manufacturers. In order to 
ensure the availability
of the beam interlock system in the future, selection of the components is now
under study. In order to take advantage of their reliability, interoperability
and the ease of insertion and extraction, SFP transceivers are being used for 
the research and development phase of the next version of the beam interlock 
system. 

The monitoring and diagnostics capabilities of these devices, combined with the
availability of manufacturers makes it an appropriate choice. In addition, the
variety of configurations that they allow is ample enough to cover many 
scenarios at the CERN accelerator complex. SFP transceivers from most 
manufacturers have versions 
with different output power values, leading to several power margins that can 
be used to accommodate the various fibre lengths of each link, or even to
compensate for increased fibre attenuation.

The BIS upgrade must be compatible with the 
current systems, meaning the interfaces and signals have to be compatible
with the original system.
Due to the compatibility requirement, the new programmable devices have to be
carefully chosen in order to be capable of implementing all input and output
signals that exist in the current BIS.

The implementation of the Beam Permit Loops will also be studied, checking the 
possibility of transmitting data messages instead of a fixed loop frequency.

\section{Conclusion}

Fibre attenuation increase due to 
radiation is a concern for the BIS, as it affects its reliability and 
availability of the LHC. The current system lacks monitoring capability and 
the addition of a measurement system is troublesome.
The use of commercially available, standardised SFP transceivers enables the
measurement of the optic fibre attenuation, either active fibres through
wavelength division multiplexing or spare fibres. 

A monitoring system has been installed in the LHC spare fibres to measure the
time evolution of fibre attenuation in high radiation areas.

SFP transceivers are also being evaluated for the next version of the 
beam interlock system, to transmit the beam 
permit loop signals at the CERN accelerator complex. The flexibility and 
features provided by these transceivers are of great interest to the next 
version of the BIS.

% \section{acknowledgment}
% Any acknowledgment should be in a separate section directly preceding
% the \SEC{References} or \SEC{Appendix} section.

%
% this setting when the default (\flushend)
% => "balance two column" shows bad results
%
% \iftrue   % balancing with bad results
% 	\newpage
	\raggedend
% \fi

% \section{appendix}
% Any appendix should be in a separate section directly preceding
% the \SEC{References} section. If there is no \SEC{References} section,
% this should be the last section of the paper.

\iffalse  % only for "biblatex"
	\newpage
	\printbibliography

% "biblatex" is not used, go the "manual" way
\else

%\begin{thebibliography}{9}   % Use for  1-9  references

\fi

\end{document}